# FROM HUMAN FACTORS TO HUMAN-TECHNOLOGY FACTORS: AN HCI PERSPECTIVE ON TECHNOLOGY IN AVALANCHE SAFETY

Björn Hartmann*[1], Jason Smith[2]

[1]*Electrical Engineering and Computer Sciences Department, UC Berkeley, CA, USA*
[2]*AMGA Ski Guide, Truckee, CA, USA*

ABSTRACT: The avalanche safety literature has identified human factors that contribute to accidents, yet researchers note a persistent gap between these insights and effective product design. Meanwhile, interactive technologies are shaping backcountry decision making with or without grounding in theory and research. We broaden the discussion of human factors into “human-technology factors,” examining how technology can both support and undermine judgment in avalanche terrain, from a human-computer interaction (HCI) perspective. We summarize relevant HCI concepts along four dimensions—attention, cognition, trust, and social interaction—and use them to revisit McCammon’s FACETS framework, cataloging ways in which specific technologies may mitigate or exacerbate classic heuristic traps, grounded in accident reports and literature where possible. We identify recurring patterns, including technologies with dual-sided effects and a pervasive “digital expert halo.”

## 1. INTRODUCTION

The avalanche safety literature has established the important role of human factors in accidents, identifying recurring patterns and connecting them to theories from the social sciences (Fredston et al. 1994; Atkins 2000; McCammon 2002; Hetland et al. 2025). Yet researchers have also pointed out a gap in translating these insights into effective products (Haegeli et al. 2023). Meanwhile, technologies are proliferating quickly, with or without rigorous grounding in theory. Interactive software and electronic devices are shaping how travelers plan, navigate, and make decisions in avalanche terrain – from the user interfaces on beacons to information visualization on forecasts and in maps, crowdsourced observations, and trip reports on social media.

This leads us to broaden the discussion of human factors in avalanche safety to “human-technology factors”: how technologies can both support and potentially undermine human decision making. We proceed from a human-computer interaction (HCI) perspective. HCI is an interdisciplinary field concerned with designing, building, and evaluating technologies that people can use effectively (Hornbæk et al. 2025). HCI grounds design in an understanding of human perceptual, cognitive, and behavioral abilities; develops interfaces through iterative design cycles; and evaluates those interfaces with their intended users to determine how well they support real needs and activities.

* *Corresponding author address:*
Björn Hartmann, EECS Department,
University of California, Berkeley, CA 94720;
email: bjoern@eecs.berkeley.edu

Technology may open new opportunities for intervention (e.g., eliciting individual assessments before group discussion in trip planning) but can also introduce new failure modes (e.g., overreliance on digital models at the expense of ground-truth observation in the field).

How might we understand these opportunities and challenges in a principled way? First, we summarize ideas from Human-Computer Interaction in four concise dimensions: Technology has effects on attention, cognition, trust, and social interaction. Second, we revisit the FACETS framework (McCammon 2004) and describe where each heuristic might be mitigated or exacerbated by technology, with pointers to accident reports, papers and articles as supporting evidence where available. We also connect these examples to the HCI dimensions. Many of these links are currently suggestive, rather than causally established in research. And in fact, sometimes attempts at establishing causal links and mechanisms have failed to find them. This creates an agenda for future research.

## 2. BACKGROUND: HCI CONCEPTS

We first summarize some key concepts and approaches in Human-Computer Interaction. The foundations of the field lie at the intersection of Psychology and Cognitive Science, Design, and Computer Science (Carroll 2003).

### 2.1. Human Capabilities And Limitations

An empirical understanding of how people perceive, attend, remember, learn, and act has given rise to design principles for appropriate interac-tive technology (Card, Moran, et al. 1983; Norman 1988). HCI also recognizes that interaction is situated within social practices, so effective designs must account not only for individual cognition but also for context, collaboration, and the distribution of knowledge across people and artifacts (Suchman 1987; Hutchins 1995).

Table 1: Potentially positive and negative effects of technology on backcountry decision-making, organized by structuring dimensions of attention, cognition, trust, and social interaction.

| Dimension | Potentially positive impact | Potentially negative impact |
|---|---|---|
| **Attention** | *Cueing:* directs attention to hazards, changing conditions, decision points, or otherwise overlooked evidence. | *Attention capture:* directs attention to digital cues at the expense of terrain, weather, partners, etc. |
| **Cognition** | *Cognitive scaffolding:* supports memory, reasoning, comparison, reflection, or the integration of distributed information. | *Cognitive offloading:* replaces reasoning with passive consumption of information, reliance on outputs without understanding the underlying structure; can lead to skill erosion. |
| **Trust** | *Calibration:* provides evidence that can counter misplaced intuitions. | *Automation bias:* leads to excessive reliance on digital information over other sources, even though it may be incomplete, stale, uncertain, or noisy. |
| **Social Interaction** | *Collective intelligence:* supports coordination, sharing of observations, collective learning, and participation in broader knowledge communities. | *Groupthink and social amplification:* exacerbates fear-of-missing-out, and improperly amplifies certain voices through selective visibility. |

### 2.2. Design

Because scientific accounts of human perception, cognition, and social behavior remain incomplete, HCI also adopts a design stance: researchers identify needs by observing and interviewing prospective users in context (Beyer and Holtzblatt 1998; Goodman et al. 2012), then iteratively prototype and refine designs through evaluation, including expert inspection, usability testing, controlled experiments, and field studies (Hornbæk et al. 2025).

### 2.3. Utility, Usability, Unintended Consequences

Utility describes whether software has the functionality needed to support a user in a given task – is it useful and fit for purpose? Usability describes whether intended users can successfully understand and use that functionality (Nielsen 1993). A system may be easy to use but offer little value, or provide powerful functionality that users cannot successfully access or control, so both must be evaluated. Even when software is both useful and usable, its widespread adoption may still produce unanticipated consequences and broader harms that are difficult to foresee during design (Nathan et al. 2008).

## 3. CONNECTING HCI CONCEPTS TO AVALANCHE SAFETY

People use a variety of interactive software and devices before, during, and after travel through avalanche terrain – some designed for that purpose by the avalanche community, others appropriated despite not being designed with it in mind. Either way, these technologies can have positive as well as negative impacts on backcountry safety. Positive impacts arise when software is explicitly designed to address a well-understood need, respects design principles grounded in empirical knowledge about people, and is evaluated with actual users to confirm benefits materialize in practice. Critical negative impacts can arise from wrong or incomplete functionality (something different from what people need, or something unreliable); usability problems (people cannot use the product as intended); inappropriate reliance on technology where it is not warranted; and software designed at cross purposes with safe backcountry practices.

We posit that technologies can be evaluated by examining their effects along four dimensions: attention, cognition, trust, and social interaction. Along each dimension, technology may support safer decisions or introduce new vulnerabilities.

The **attention** dimension asks: What does technology make salient, and what does it obscure? Technology can direct attention to consequential cues, such as recent avalanche activity, but can also divert attention from changing field conditions or create attentional tunneling.

The **cognition** dimension asks: What forms of reasoning does technology support, and what does it replace? Information visualization can enable "using vision to think" (Card, Mackinlay, et al. 1999), while automated recommendations can discourage critical reflection.

The **trust** dimension asks: How does technology shape judgments of credibility and authority? It can help users identify reliable, well-supported information, but polished presentation or apparent expertise can also engender unwarranted trust.

The **social interaction** dimension asks: How does technology reshape communication, coordination, and social influence? It can support shared awareness and constructive dissent, but can also amplify social proof, suppress disagreement, or promote groupthink.

We summarize the nature of potential positive and negative effects in Table 1.

## 4. REVISITING FACETS

To gain another perspective of potential technology impacts, we revisit the FACETS framework (McCammon 2002; McCammon 2004). FACETS identified a memorable set of recurring heuristic traps that can distort judgment in avalanche terrain: familiarity, acceptance, commitment, expert halo, tracks/scarcity and social facilitation. While the heuristic traps are universal (because they are about people), adding interactive technology into the mix has the potential to either mitigate or exacerbate these traps.

### 4.1. A Catalog Of Technology Impacts

We describe 16 potential impacts of technologies on FACETS in Tables 2 and 3. We ground our analysis in references to public reports, articles and research literature where possible but acknowledge that some of these links are speculative and not proven. In addition, the table is not exhaustive — there are additional positive impacts where technologies work as intended, and additional negative impacts we failed to envision. The technology landscape and usage patterns also constantly evolve, so this is a starting point and a snapshot rather than an encyclopedic list.

Each row in the table contains a summary of the impact, the most salient FACET, the technology that causes the impact, whether the impact affects attention, cognition, trust or social interaction, a source or example, and a judgment whether the impact is likely positive, negative, or both.

We further distill these examples into a more memorable format that could be used in instruction in Figure 1.

### 4.2. Larger Patterns

The following larger patterns emerged from our work. First, many technologies can have both positive and negative impacts on individual heuristic traps. For example, social media can mitigate the acceptance trap if they are used as a form of accountability. Sometimes people share close calls or bad decision making. Or commenters call out and discuss questionable decisions in a post. On the other hand, social media can exacerbate the very same heuristic when posters feel pressure to produce content during elevated danger conditions for their followers. In addition, technologies may have negative impacts in one heuristic but positive in another one. For example, map layers can mitigate familiarity by revealing additional information but can also introduce a digital expert halo when they show information that looks authoritative but may have important limitations. These two-sided impacts mean that there is no simple message to avoid certain technology use. Rather, users will have to learn to be discerning about when and when not to use or trust a particular technology. Thus it is important for avalanche education to talk about these impacts.

Second, many of the impacts we cataloged concern the topic of “digital expert halos.” Decision making in complex situations is cognitively demanding and there are many uncertainties. Digital information often promises to transcend that uncertainty by offering clear, unambiguous, authoritative and reliable data. But just because something is rendered on a screen does not mean our trust is warranted. Weather sensors on mountain tops can stop functioning; the limited resolution of maps can hide important terrain features such as smaller cliffs; AI chatbots don’t take changing conditions into account when recommending routes; real avalanches may behave differently than runout simulations in map layers suggest; a low danger rating does not mean we are free to stop thinking when we venture into terrain. Data presented in apps *can* be more objective than personal opinion and intuition, but users need to understand the limitations of its source, accuracy, and applicability.

Finally, social pressures (through tracks/scarcity and social facilitation) don’t just apply through direct communication between people — they can also appear asynchronously, over time, mediated through data. Heatmaps that aggregate collective behavior over time are a prime example, and it’s likely others will emerge as we aggregate, process, and re-visualize additional types of data in our digital platforms in the future.

## 5. RELATED WORK

We briefly review prior work published at ISSW and related venues that is of relevance.

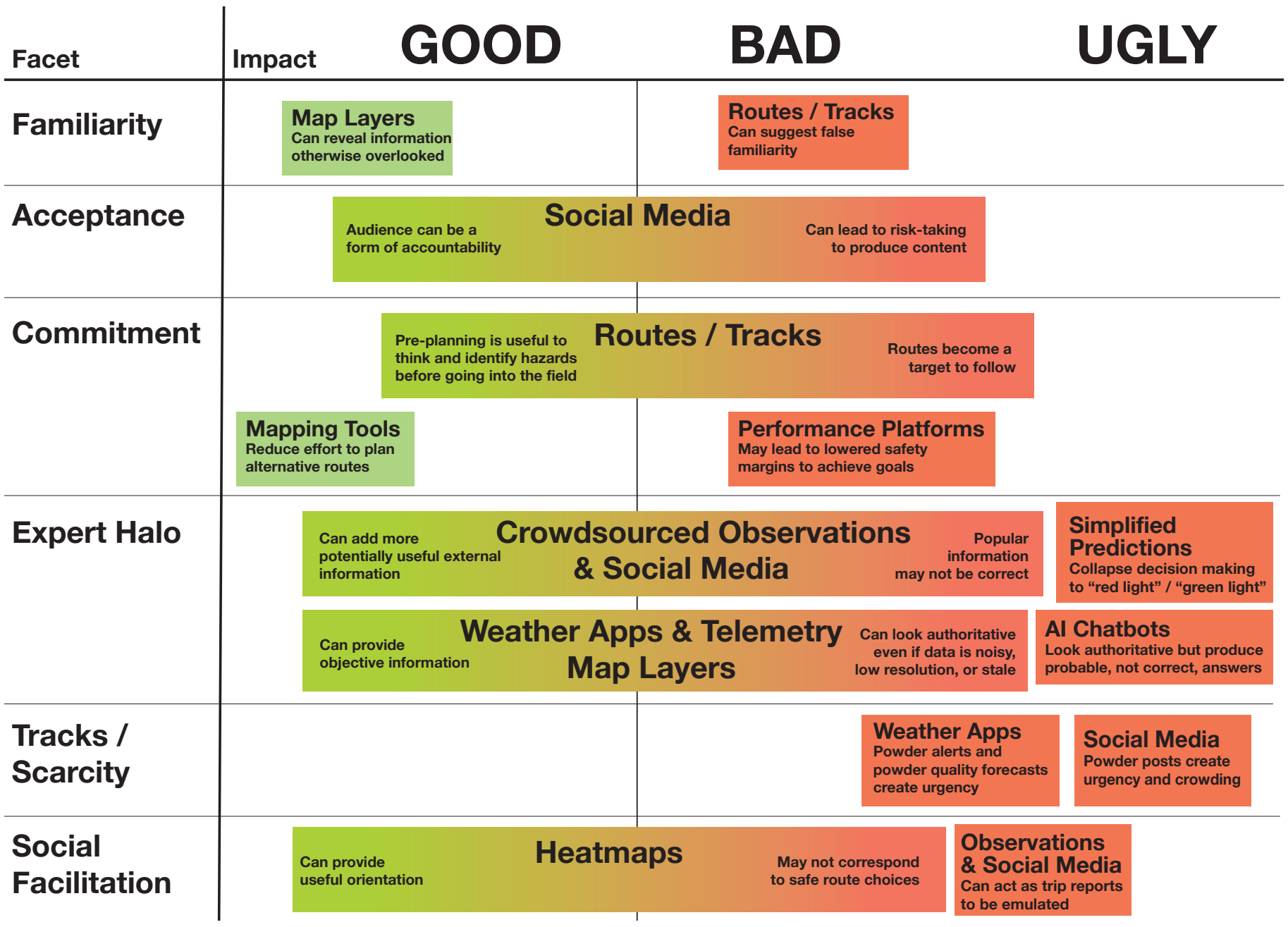

Figure 1: Revisiting FACETS: technologies can either mitigate or exacerbate existing heuristic traps.

### 5.1. *Human Factors*

Several seminal works introduced ideas from cognitive science and the social sciences around biases to avalanche safety (Fredston et al. 1994; Atkins 2000; McCammon 2002). More recently, Hetland et al. produced a scoping review of the literature around human factors (Hetland et al. 2025). However, Haegeli and others note that this awareness of social science literature has not necessarily led to better solutions (Haegeli et al. 2023).

### 5.2. *Interactive Technologies at ISSW*

Many ISSW papers describe novel user interfaces aimed to help certain groups of users, for example (Ruesch et al. 2013; Bourova et al. 2014; Vidaud-Barral et al. 2010; Proksch et al. 2018; Cremonini et al. 2013). The papers at ISSW often focus on features or technical implementation, less frequently on evaluations with users that could determine impacts in practice. An interview study with creators of six different avalanche applications finds that evaluation of the applications was an underdeveloped aspect (Charrière and Bogaard 2016). In contrast, the work of Isaak stands out as it starts from human factors and then extends those concepts to social media (Isaak 2016; Isaak and Sather 2023).

### 5.3. *The Intersection of HCI and ISSW*

Some previous HCI researchers have had connections with ISSW, making contributions to both snow science and their respective home communities. For example, Desjardins' research on beacon parks appears both at ISSW and at CSCW (Desjardins et al. 2016a; Desjardins et al. 2016b). Similarly, Nowak's work on data visualizations appears both at ISSW (Nowak, Haegeli, et al. 2023) and in the Visualization conference (Nowak and Bartram 2024). However, overall this crossover is rare.

In summary, while the ISSW community is very attentive to the presence of human factors in decision making at a conceptual level, and while it has developed many applications to communicate important information that is useful for decision making, how the design of software or devices interacts with human factors, and if it ultimately addresses them has received less attention.

## 6. CONCLUSION AND FUTURE WORK

We have described potential positive and negative impacts of using interactive technologies before, during, and after travel in avalanche terrain. Our work is a starting point, but not a full exploration. While some impacts have been documented in accident reports, articles and the research literature, their prevalence and relative importance is often still unknown or speculative. Impacts will change over time as technologies and people's usage patterns evolve. Our judgment of impacts is also subjective. We invite the community to give feedback on this project, and suggest other technology impacts, at `techfacets.org`.

Table 2: An account of how digital tools interact with human factors in avalanche decision-making (1 of 2).

| Summary | FACET | Technology | ACTS | Source / Example | G/B/U |
|---|---|---|---|---|---|
| Map overlays can mitigate familiarity by directing attention to hazards that are otherwise overlooked. | Familiarity | Mapping and Navigation | Attention | Example: In a route-planning exercise, an avalanche educator mapped a familiar tour commonly regarded as safe. Adding digital slope-angle shading revealed that the standard route passed beneath and potentially through several avalanche paths (Diamond 2023). | **Good** |
| Digital routes of previous trips can exacerbate familiarity by suggesting a route is safe. | Familiarity | Routes and GPX tracks | Attention, Trust | Example: Skiers relied on a pre-existing GPX track in Norway that was possibly created for summer travel, assumed it encoded a sub-30° winter route, and triggered an avalanche (neuvilla 2025). | **Bad** |
| Social media can either mitigate or exacerbate acceptance by exerting ”invisible pressure” through a virtual audience. Good pressure can manifest as accountability, bad pressure can manifest as a need to deliver more content, even under unsafe circumstances. | Acceptance | Social Media | Social Interaction | Papers: Isaak posits “invisible pressure” as a factor that exacerbates acceptance (Isaak 2016). One study of adolescents found a positive correlation between social media use and risk taking, while another one did not (Cornell and Peden 2025; Frühauf et al. 2025). | **Good** or **Bad** |
| Digital mapping tools can mitigate commitment by lowering the effort required to generate and compare alternative routes. | Commitment | Mapping and Navigation | Cognition | Paper: Kosberg studied decision making with augmented map layers (Kosberg et al. 2024). | **Good** |
| Pre-planning routes can either mitigate or exacerbate commitment. They can mitigate by allowing users to reason about and identify hazards ahead of time, before they need to make decisions in real time in the field. They can exacerbate commitment when the route becomes a target to complete, potentially in the face of contradictory evidence. | Commitment | Routes and GPX tracks | Attention, Trust | Paper: “Death by GPS” occurs in navigation in cars (A. Y. Lin et al. 2017). | **Good** or **Bad** |
| Performance tracking apps can exacerbate commitment by making speed or records the dominant objective, at the cost of reduced safety margins—e.g., including carrying less emergency equipment. | Commitment | Activity and performance tracking platforms | Attention | Example: Winter FKT attempt with minimal gear; the athlete acknowledges this “decreases the margin of error” (Maune 2011). | **Bad** |
| Crowdsourced observations can either mitigate expert halo by adding outside perspectives, or introduce new expert halos when they are incorrectly seen to be authoritative (come from popular/ famous sources) . | Expert Halo | Crowdsourced Observation | Social Interaction | Paper: Crowdsourced observations are useful, including for forecasters (Tremper and Diegel 2014).<br>Paper: On the other hand, people use heuristics to judge source credibility in online media that are independent of content (X. Lin et al. 2016). | **Good** or **Bad** |
| Digital map layers can provide objective information or create a digital expert halo by presenting data with an appearance of precision and authority, even if the underlying data sources have limited resolution, are noisy, or are derived from simulation with limited accuracy. | Expert Halo | Mapping and Navigation | Trust | Example: In the 2019 Silverton Avalanche School accident, the group used slope-angle shading to identify what appeared to be a sub-30° descent; the actual slope measured 32–34° (O’Neil 2022).<br>Example: In the 2025 Radio Tower Peak accident, the group consulted mapping apps and selected what appeared to be a low-angle traverse beneath steeper terrain. Investigators noted that commonly used 10-meter digital elevation models can entirely omit a cliff smaller than one grid cell. (Bridger-Teton Avalanche Center 2025). | **Good** or **Bad** |
| Weather forecasts and telemetry can provide objective information or introduce a digital expert halo, suggesting certainty that isn’t warranted when actual weather locally diverges from forecast points or when instruments aren’t calibrated or working. | Expert Halo | Weather telemetry and forecasts | Trust | Example: Actual weather events can exceed forecasts (Avalanche Canada 2020).<br>Example: Telemetry sensors may not be in the right locations to provide relevant info (Missoula Avalanche Center 2024). | **Bad** |

Table 3: An account of how digital tools interact with human factors in avalanche decision-making (2 of 2).

| Summary | FACET | Technology | ACTS | Source / Example | G/B/U |
|---|---|---|---|---|---|
| Social media can either mitigate or exacerbate expert halo in ways similar to crowdsourced observations, but with even noisier data and worse source credibility problems. Accounts with high algorithmic reach might be seen as authoritative, even if they are not experts. | Expert Halo | Social Media | Social Interaction | Paper: People use heuristics to judge source credibility in online media that are independent of content (X. Lin et al. 2016). | **Good** or **Bad** |
| Apps that reduce nuanced conditions to red-light/green-light recommendations can create an algorithmic expert halo that discourages critical thinking. | Expert Halo | AI Apps | Attention, Cognition | Novices tend to rely more heavily on the danger rating to make trip planning decisions (Morgan et al. 2023).<br>Green danger ratings can be conflated with a "green light" (Conger 2004). | **Ugly** |
| AI applications like chatbots can introduce a digital expert halo when they produce recommendations that look authoritative but are not tailored to present time conditions. | Expert Halo | AI Apps | Trust | Example: Lions Bay SAR warns against using ChatGPT after a rescue (Ghuman 2025).<br>Example: Hikers rescued from Mount Shasta after using AI to plan trip (Sweeney 2026) | **Ugly** |
| Powder alerts in weather apps can exacerbate tracks/scarcity by creating urgency. | Tracks/ Scarcity | Weather telemetry and forecasts | Cognition | Example: Apps deliver "powder alerts" and "powder quality forecasts" (OpenSnow 2024; OpenSnow 2025). | **Ugly** |
| Social media can exacerbate tracks/ scarcity: seeing powder skiing content from a specific zone creates urgency and potentially leads to crowding which in turn can also lead some to push past safer, heavier traveled terrain. | Tracks/ Scarcity | Social Media | Social Interaction | Example: Vermont rescue team sees an increase in calls as social media lures skiers into danger (Swinhart 2025). | **Ugly** |
| Crowdsourced observations can exacerbate social facilitation when they act as successful trip reports - they can provide social proof that others successfully skied similar objectives, potentially encouraging emulation. | Social Facilitation | Crowdsourced Observations and Social Media | Social Interaction | Paper: Posts showing other parties completing an objective may provide a model for others to emulate (Isaak 2016); non-events can be misconstrued as evidence that the terrain choice was sound. | **Bad** |
| Heatmaps can potentially exacerbate social facilitation. While they can show useful common travel corridors that can aid orientation, heatmaps aggregate across time, and routes may not be safe under current conditions. | Social Facilitation | Activity and performance tracking platforms | Trust, Social Interaction | Article: Dangers and opportunities of heatmaps (in German) (Prantl 2024). | **Good** or **Bad** |